# Direct Laser Micropatterning of GeSe$_2$ Nanostructures with Controlled Optoelectrical Properties

Bablu Mukherjee[1], Govinda Murali[2], Sharon Xiaodai Lim[1,3], Minrui Zheng[1], Eng Soon Tok[1], and Chorng Haur Sow[1*]

[1]Department of Physics, 2 Science Drive 3, National University of Singapore, Singapore-117542

[2]Department of Physics & Meteorology, Indian Institute of Technology Kharagpur, India-721302

[3]Graduate School of Integrative Sciences and Engineering, National University of Singapore, 28 Medical Drive, 117456, Singapore

ABSTRACT

We demonstrate that a direct focused laser beam irradiation is able to achieve localized modification on GeSe$_2$ nanostructures (NSs) film. Using scanning focused laser beam setup, micropatterns on GeSe$_2$ NSs film are created directly on the substrate. Controlled structural and chemical changes of the NSs are achieved by varying laser power and treatment environment. The laser modified GeSe$_2$ NSs exhibit distinct optical, electrical and optoelectrical properties. Detailed characterization is carried out and the possible mechanisms for the laser induced changes are discussed. The laser modified NSs film shows superior photoconductivity properties as compared to the pristine nanostructure film. The construction of micropatterns with improved functionality could prove to be useful in miniature optoelectrical devices.

## INTRODUCTION

Focused laser system has found great utility in nanoscience research. It has a wide variety of applications ranging from micropatterning, modification of nanomaterials, manipulation of nanorods and creation of nanorotors using optical traps.[1-3] Scientists have recently been able to produce laser beams of size as small as that of a virus which could further enhance the already proven ability of laser to create amazing 3-D structures in nanoarrays.[4-6] Focused laser modification has been extensively employed in different materials including carbon nanotube,[7] graphene oxide,[8] CuO nanorods etc.[1] Nanomaterials modified with focused laser beam often show varied and tunable structural, optical as well as electrical properties. Focused laser beam irradiation with the sample housed in controlled gas environments further enhances the degree of variation in experimental condition during the modification of the materials and could have many useful applications.[9] It has been shown that focused laser beam irradiation is able to convert less conductive graphene oxide to more conductive reduced form.[10] Recently Lu *et al.* have shown multifunctional micropatterns fabrication on $CdS_xSe_{1-x}$ nanobelt films using scanning focused laser beam where the samples exhibit modification of chemical composition. And the applications of such modified nanobelts as a potential acid sensor and more superior photoconductor have been demonstrated.[11] Considering its simplicity, efficiency and flexibility, nanomaterial modification with focused laser system thus appears to be an attractive approach in nanomaterial research. Furthermore, direct patterning of the nanomaterials using focused laser beam has an added advantage when compared with other patterning methods such as nanoimprinting, photolithography, microcontact printing etc. because of minimized chemical contamination to the sample.[12-14]

Germanium diselenide ($GeSe_2$) has an interesting crystallization process which can be controlled using light irradiation or thermal-annealing.[15] The photo-induced and thermal crystallization processes of amorphous $GeSe_2$ films have been well studied using Raman spectroscopy. Thermally annealed amorphous $GeSe_2$ shows the low temperature form (LT-$GeSe_2$) of crystalline $GeSe_2$, On the other hand if the amorphous $GeSe_2$ film is thermally annealed at a higher temperature for longer duration, then amorphous $GeSe_2$ turns into high temperature form (HT-$GeSe_2$) of crystalline $GeSe_2$.[16] Similarly amorphous $GeSe_2$ crystallized into HT phase rather than LT phase due to the thermal heating and photo-induced effect caused by the laser absorption.[17] Earlier studies on $GeSe_2$ focused on the properties of $GeSe_2$ films. Recently single-crystalline germanium diselenide ($GeSe_2$) nanostructures (NSs) have become promising candidates for field emission devices,[18] supercapacitor and photodetector devices.[19,20] Single crystalline $GeSe_2$ has anisotropic thermal and electronic properties.[21]

Chalcogenides glasses are interesting due to their potential in optical and photonic applications.[22] In the past decades, a large number of studies including the structure of glassy $GeSe_2$ have been reported.[23,24] Glassy $GeSe_2$ forms good covalent bonded glass. With the presence of copper, the local structure of glassy $GeSe_2$ is changed and it exhibits photo-darkening.[25] Photo-induced change of optical properties has been studied on glassy $GeSe_2$ films, which shows the change in optical transparency of the films.[26] The photo-induced local structural changes have been observed in GeSe$_2$ films using in situ extended x-ray absorption fine structure (EXAFS) analysis.[27] Hence glassy GeSe$_2$ attracted attention as a candidate that exhibits photo-induced changes in properties. However, till now there are no such reports on crystalline GeSe$_2$ NSs. Being a class of interesting multifunctional semiconductors with layer structure, single crystalline GeSe$_2$ NSs in the form of nanobelts show interesting structural changes during laser modification with different laser powers. Most recently, we reported the synthesis of stepped morphologies GeSe$_2$ nanobelts, where the photosensing properties of individual GeSe$_2$ nanobelt were characterized using light excitations with different wavelength.[20] One of the possible reasons for the good photodetecting property was found to be the presence of defects associated with the individual nanobelt device.

In this work, high quality single crystalline GeSe$_2$ NSs were synthesized through Au-catalyzed one-step VLS process using horizontal tube furnace. Focused laser micro-modification of these NSs was demonstrated. The focused laser beam was found to facilitate structural changes in these NSs as well as changes in their optoelectrical properties. The controlled structural changes were investigated on crystalline GeSe$_2$ NSs film using Raman spectroscopy. Multicolored micropatterns were created on GeSe$_2$ NSs film under controlled gas environment in air, vacuum and helium. For the investigation of potential device application, two electrical contacts were connected directly on the uniform NSs film in an easy and efficient way. The photoconducting properties of the nanonetwork device with pristine and modified NSs film were studied with different laser irradiation. Remarkably we found that laser modified NSs exhibit more superior photoconducting properties.

## RESULTS AND DISCUSSION

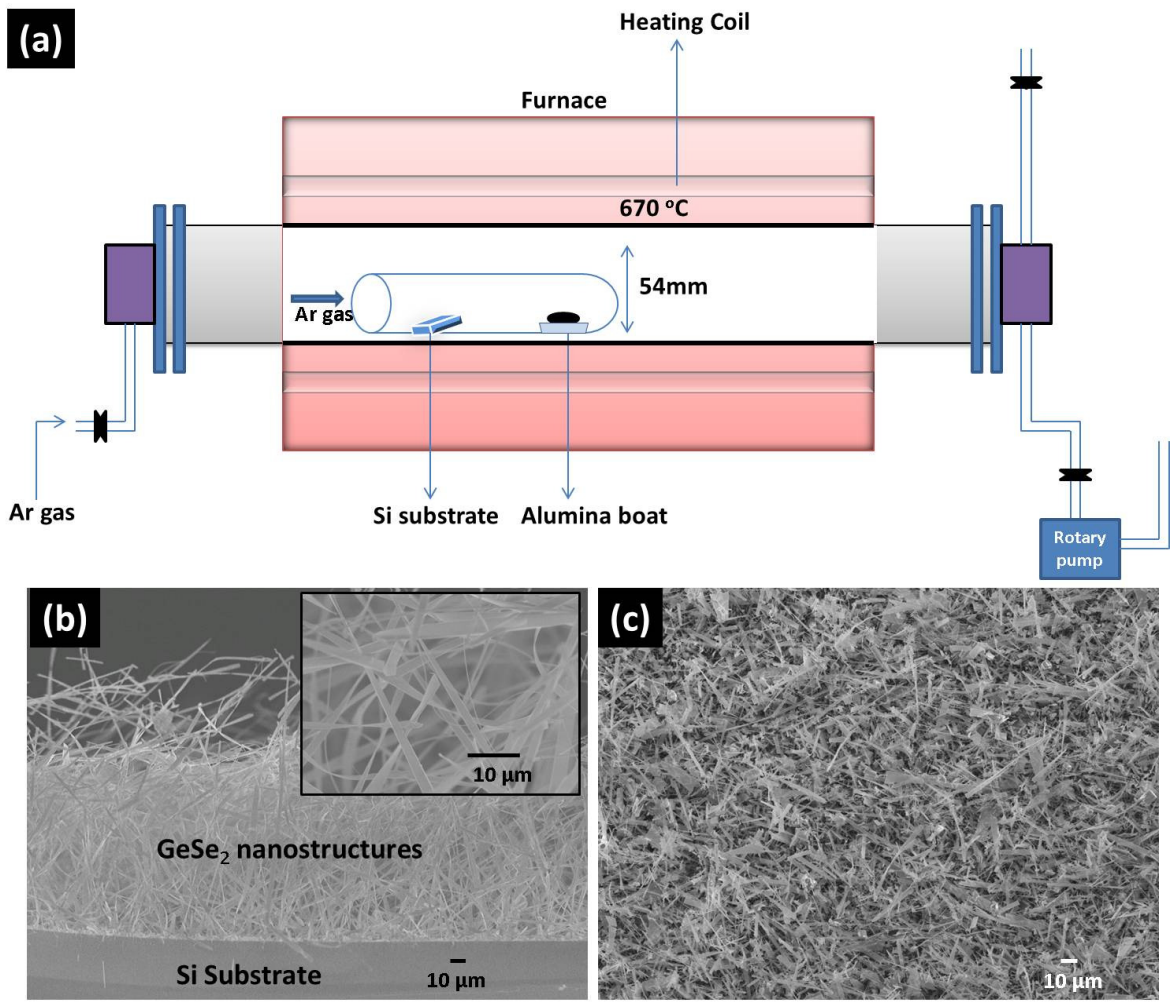

**Figure 1 (a)** Schematic illustration of the experimental setup for the synthesis of GeSe$_2$ NSs. **(b)** SEM image of the cross-sectional view of as synthesized NSs on Si substrate. Inset shows the HRSEM image of the top-view of the product. **(c)** SEM image of the NSs after mechanical pressing.

GeSe$_2$ NSs were synthesized in a horizontal furnace via a chemical vapor deposition (CVD) process as described in the experimental section. The conditions and schematic representation of the synthesis set-up is shown in **Figure 1(a)**. Upon completion of the synthesis, yellow colored GeSe$_2$ NSs were found on Si substrate. Cross-sectional SEM image of as grown GeSe$_2$ NSs on Si substrate before mechanical pressing is shown in **Figure 1(b)**. Evidently, high density of NSs was grown on the substrate with the average length of nanorods being more than 80 µm. A top-view higher resolution SEM (HRSEM) image of as grown GeSe$_2$ NSs is shown in the inset of **Figure 1(b)**. SEM images show that the nanobelts with an average width of ~ 1-2 µm and a length of ~ 50-100 µm were obtained during synthesis. The NSs were then pressed with a glass slide to obtain more compact and uniform NSs on the substrate. **Figure 1(c)** shows SEM image of the pressed GeSe$_2$ NSs on Si substrate. The density and uniformity of the NSs is further improved by mechanical pressing.

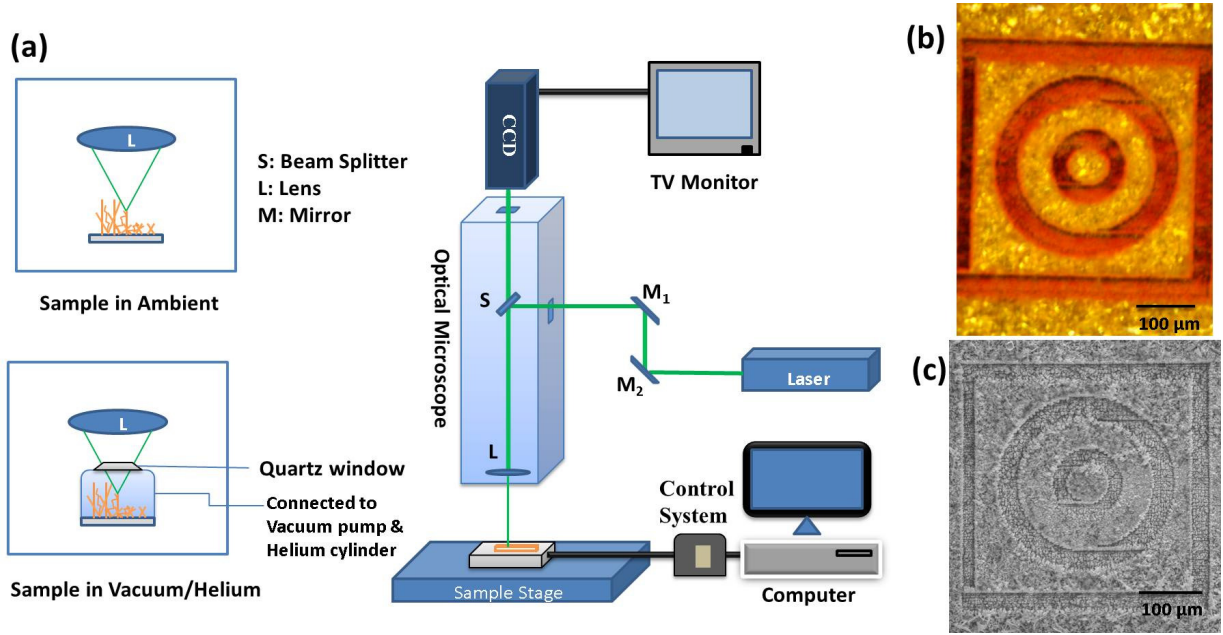

**Figure 2 (a)** Schematic representation of the optical-microscope with focused laser beam setup for micropatterning. **(b)** Optical microscope and **(c)** SEM images of laser patterned circular rings in a square on GeSe$_2$ NSs film.

Direct focused laser micropatterning and micromodification were carried out through a typical optical microscope system that was coupled with an external laser beam as shown in **Figure 2(a)**. A SUWTECH LDC-2500 diode laser ($\lambda$ = 532 nm, maximum power ~ 300 mW) was used for our focused laser beam setup. The parallel emitted laser beam was then directed into an upright optical microscope via two reflecting mirrors (M$_1$ and M$_2$). Inside the microscope, the beam was directed towards an objective lens via a beam splitter. The laser beam was then focused by the objective lens with a 50× magnification, a numerical aperture of 0.55 and a long working distance of 8.7 mm. The size of the focused laser beam was about ~ 4-5 μm in diameter. The samples were placed on a computer-controlled, motorized stage (MICOS VT-80 System) with a minimum step size of 1 μm in the *x*–*y* direction. By moving the sample stage in a controlled programmable manner with respect to the focused beam, we were able to create a wide variety of micropatterns on the substrate. The same objective lens (L) was used to collect light reflected from the sample for viewing purposes. A CCD camera was used to capture the images of the laser trimming process. Hence during the laser modification of the NSs, we could visually inspect the structures created simultaneously through a TV monitor. Apart from ambient conditions, the micropatterning was carried out in vacuum or Helium environment with the sample housed in a vacuum chamber (dimensions: L×W×H ~ 2cm × 1cm × 0.5cm) connected to a vacuum pump and Helium gas cylinder. In this case the laser was focused towards the sample through the quartz window of the vacuum chamber.

Upon focused laser beam irradiation onto the nanostructured thin film, there would be a rapid local temperature rise due to the absorbed laser energy, which caused local heat generation. The local temperature at the center of the laser beam was directly related to the intensity of the beam. The localized intense heat eventually caused different degrees of modification and pruning on the NSs. **Figure 2(b)** shows the optical image of the pattern created using the focused laser beam setup via a laser power of ~ 2 mW and **Figure 2(c)** shows SEM image of the same pattern. Using the laser pruning method, micropatterns of circular rings in a square with different color contrasts were readily created on the film of yellow colored pristine $GeSe_2$ NSs. The color can be readily regulated and further altered by carefully controlling the laser power. More patterns were created using the same laser power (~ 2 mW), some of which are illustrated in **Figure 3**. The SEM images (**Figure 3a,c**) of the two complex microstructures created show the distinct morphological contrast of the laser pruned area due to the relatively high laser power used for modification. The optical images (**Figure 3b,d**) clearly show the distinct color contrast of the patterns, where the micropatterns are more visible.

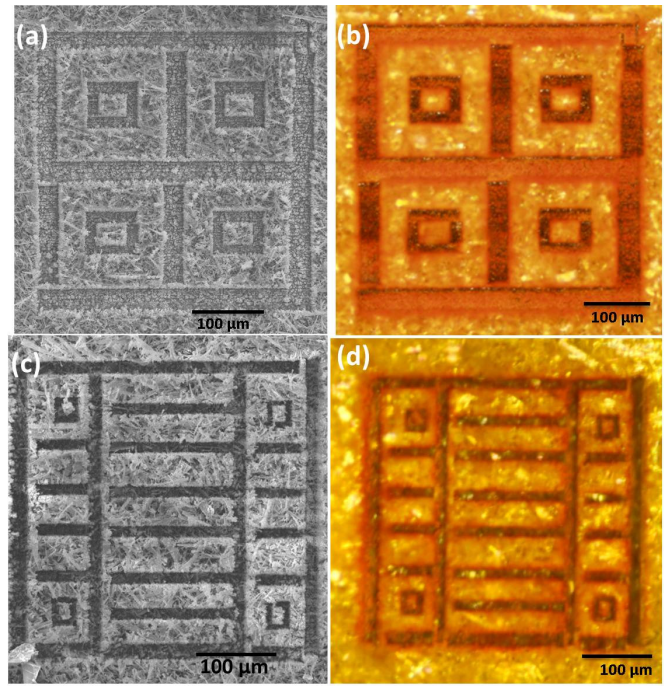

**Figure 3** **(a)** SEM, **(b)** optical microscope image of four-square patterns on $GeSe_2$ NSs film created via a focused laser beam (Wavelength: 532 nm at fixed power of ~ 2 mW). **(c)** SEM, and **(d)** optical microscope image of a micropatterned barcode created with same conditions as the four-square pattern.

The laser treated NSs had been transformed or modified into different type of materials. To understand more about the modification process, power dependent systematic studies were carried out. Four different microsquares (200 μm × 200 μm) were created on GeSe$_2$ NSs film *via* raster scan of the tiny focused laser beam with different laser powers. The laser power was increased from (i) ~ 0.2 mW, (ii) ~ 0.6 mW, (iii) ~ 1.8 mW and (iv) ~ 40 mW (power measured after focused) in four steps and the corresponding optical microscopy images of the patterns are shown in **Figure 4(a)**. The corresponding SEM images are shown in **Figure 4(b)**, where the micropatterned boxes are highlighted by white dotted lines. It is interesting to note that the box region modified with lowest laser power ~ 0.2 mW (labeled as i) was barely observable from the SEM image, which clearly indicates that the morphology of the pristine GeSe$_2$ NSs remained intact after weak (~ 0.2 mW) laser modification. With increase in laser power, the SEM images show very significant transformation in the NSs. After the irradiation of focused beam with high laser powers, the nanobelt structure of the material is completely lost and is transformed into blistered form exhibiting amorphous nature. Laser power dependent color and morphology modification of the NSs suggest that the laser induced thermal effect plays a significant role in the observed phenomena. We made use of finite element modeling (FEM) to determine the probable local temperature rises on the surface of the nanostructure film under focused laser beam irradiation. A 3D model of GeSe$_2$ NSs film (5 mm × 5mm × 50 μm) was built and subjected to allow under laser irradiation (along –z direction) of 532 nm wavelength (Gaussian beam with 3μm spot size in diameter) while the sample is moving at speed of 200 μm/s. **Figure S1** illustrates the simulated results of approximate value of the local temperature produced by different focused laser powers. The temperature distribution with different laser power is shown in Figure S1 when the laser beam has been illuminated for 1s. A volume factor of 0.4 is used to specify the porous media of the modeled NSs film with stationary air presence in the pores. Different temperature rises (**Figure S1(a-d)**) have been observed at the laser heated spot, which predicts that laser modification would have been taking place under those conditions. A maximum temperature of 1025 K has been obtained under focused laser beam with 40 mW laser power. The local temperature on the top surface of GeSe$_2$ NSs film can rise upto ~ 77 $^0$C, 190 $^0$C, 461 $^0$C and 752 $^0$C under focused laser irradiation with 0.2 mW, 0.6 mW, 1.8 mW and 40 mW, respectively. Thus it is highly probable that the localized heat due to the focused laser beam modifies the crystalline order and chemical composition of the NSs.

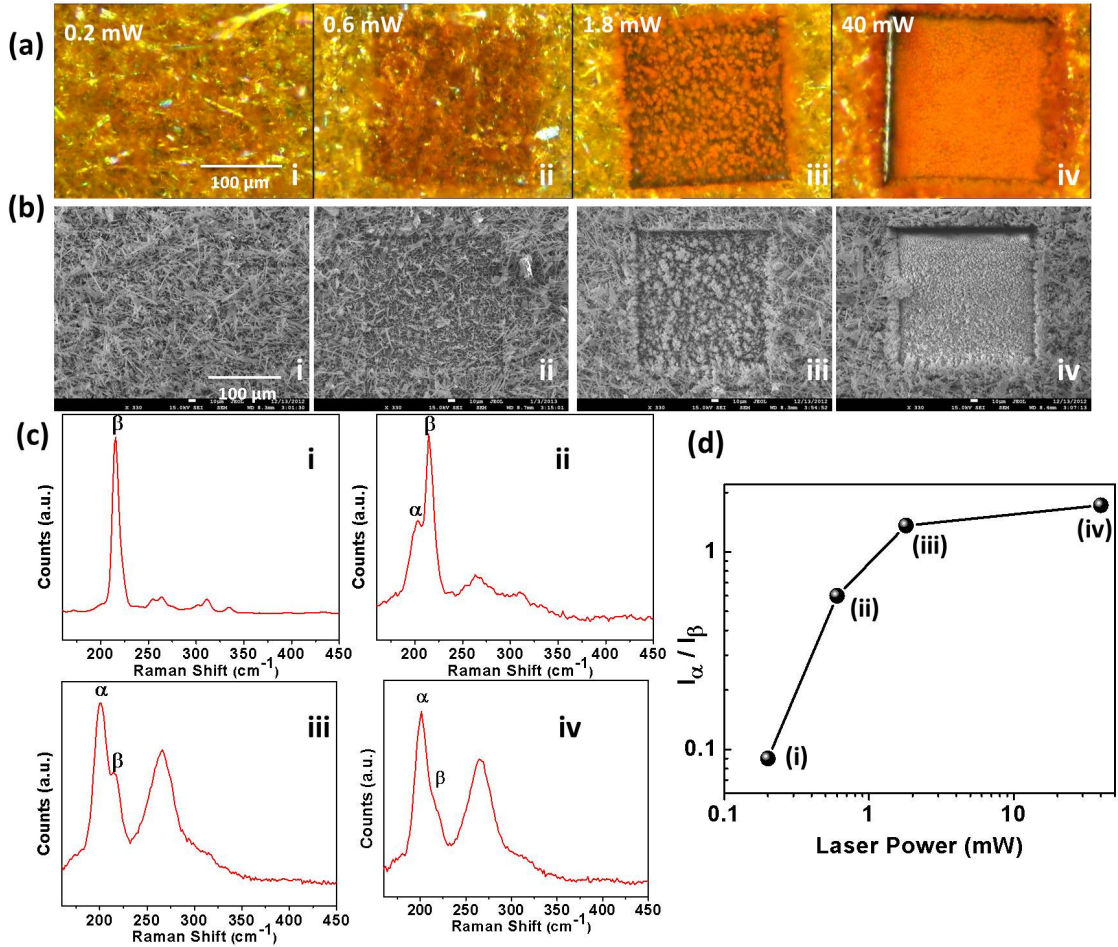

**Figure 4 (a)** Optical microscopy images, and **(b)** SEM images of four microsquares patterned on GeSe$_2$ NSs film via a focused laser beam with different laser power as indicated in the images labeled as (i), (ii), (iii), and (iv). **(c)** Raman spectra of the representative microstructures. (d) Ratio of intensity of α crystalline (I$_α$) to that of β crystalline (I$_β$) as a function of laser power. Both axes are in log-log scale.

To gain a better insight into the changes made during the laser modification process, we have obtained Raman spectra at patterned boxes modified with different laser powers. Raman spectra for each modified regions are shown in **Figure 4(c)**, labeled according to the laser power used as i to iv. The variation of the intensity ratio of α band to β band of GeSe$_2$ with laser power is presented in **Figure 4(d)**. As evident in the Raman spectra (**Figure 4(c)**), the band α at ~ 200 cm$^{-1}$ arises from the breathing mode of corner sharing tetrahedra (CST), whereas the band β at ~ 216 cm$^{-1}$ arises from the breathing mode of edge sharing tetrahedra (EST).[27] The intensity ratio of the α band to β band in the modified samples is also found to be changing with the laser power indicating that the chemical composition of the material is modified. A new broad peak around ~ 267 cm$^{-1}$ has appeared in the laser modified samples with power 1.8 mW and 40 mW, which can be seen in the Raman spectra, **Figure 4c (iii, iv)**. The broad peak located at ~ 267 cm$^{-1}$ is related to Se-Se pairs, amorphous Ge and corresponds to the presence of elemental selenium in the amorphous state.[28-30] Thermal-induced decomposition of the elements and the formation of amorphous like materials on the surface of the laser modified nanobelt could be the reason for the appearance of broad peak.

GeSe$_2$ exists in two crystallographic modifications in normal pressure conditions, low temperature (LT, α-crystalline phase) and high temperature (HT, β-crystalline phase).[31,32] The HT phase has two dimensional layered structure in which GeSe$_{4/2}$ tetrahedrons are mutually connected both in corners (corner sharing tetrahedron or CST) and edges (edge sharing tetrahedron or EST) with equal number of CSTs and ESTs.[33] The LT phase has a three dimensional structure consisting of the same GeSe$_4$ framework but connected only by the CSTs with the tetrahedron chains aligned along directions [001] and [010].[34] Apart from these two phase modifications, the amorphous GeSe$_2$ consists of the tetrahedral units connected in different CST and EST patterns. When the intensity of the laser is increased, the HT form initially transforms to LT form, as evident from **Figure 4(c,d)**, and further to a disordered amorphous form under high laser powers. Minaev *et al.* observed that during the transition in which single crystal HT phase loses its crystal structure completely and transforms to disordered non-crystalline materials, the color of the material changed from yellow to red.[35] In our case, the change in color and the corresponding change in structure as revealed by Raman spectra are consistent with their observation. Hence one of the contributing factors to the laser induced color change can be attributed to change in the state of crystallographic modification in GeSe$_2$.

HRSEM images of modified and unmodified regions of GeSe$_2$ NSs are shown in **Figure 5(a)**, which clearly shows the morphological changes of pristine NSs after laser modification with different powers. **Figure 5(b)** shows an example of the morphology of unmodified NSs. **Figure 5(c)** shows HRSEM images of NSs after laser modification with different laser powers. They show the varying degree of morphological changes of the NSs. The NSs undergo significant changes with irregular abrasive grains with a high degree of interconnected porosity and roughness. The morphology caused by high power laser-induced damage indicates the formation of the tiny and irregular grains with multi-facets.

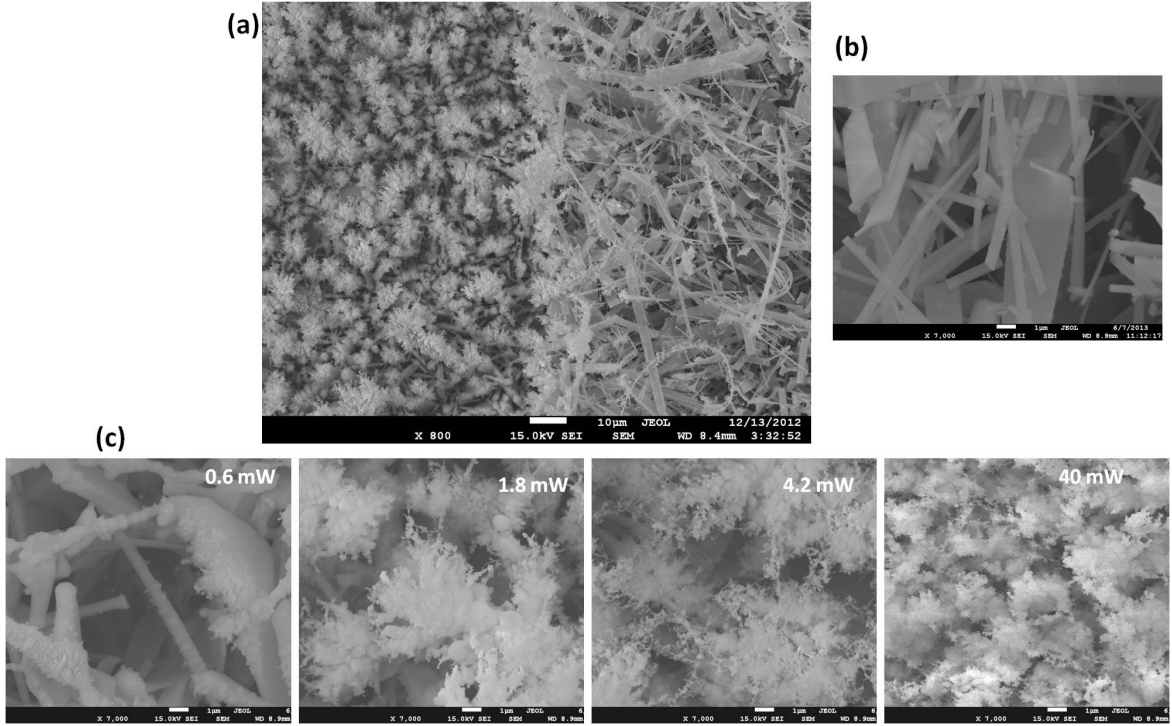

**Figure 5 (a)** HRSEM image of GeSe$_2$ NSs on Si substrate with laser modified and unmodified region. The boundary between modified and unmodified region can be clearly seen from the image. **(b)** SEM image of the pristine GeSe$_2$ NSs. **(c)** SEM images of the laser modified GeSe$_2$ NSs with different laser powers. Scale bars in **(b,c)** = 1μm.

Besides the interesting structural changes, we can use this technique to create 3D microstructures as well. The optical image (**Figure S2a**) shows the channels cut in ambient environment with the focused laser beam of power ~ 6 mW. The SEM image of the array of microchannels (separation ~ 10 μm) is shown in the inset of **Figure S2(a)**, which indicates that high spatial resolution is achievable using this direct pruning technique. The zoom-in magnified SEM image as shown in **Figure S2(b)** states that well-defined line channel with a width of ~ 1.3 μm is obtained. This channel width is smaller than the focused spot size because of Gaussian intensity profile of the laser beam. Cross sectional SEM image of a V shaped channel formation on nanostructure film is shown in **Figure S2(c)**, when focused laser with stronger laser power (~ 40 mW) has fully burned the NSs and eventually produced a V shaped channel due to profile of the focused laser spot. Conversely for the square modified with focused laser beam at 0.6 mW, the length of the NSs becomes shorter as revealed in the cross sectional SEM image (**Figure S2(d)**).

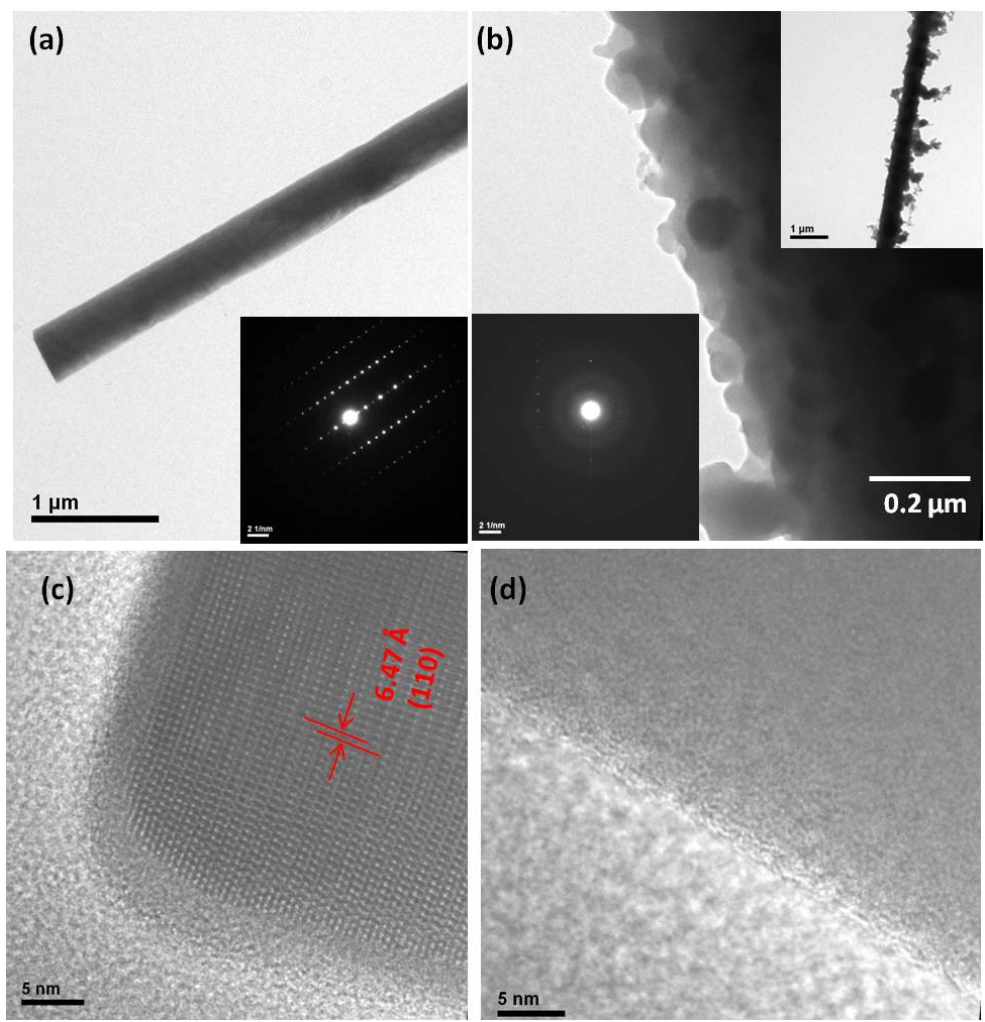

**Figure 6 (a)** TEM images of a pristine GeSe$_2$ nanobelt. The inset is the SAED pattern for the representative nanobelt. **(b)** TEM image of a laser modified GeSe$_2$ nanobelt, where the low magnified TEM image of the laser modified nanobelt is shown in the inset (top-right). The inset (bottom-left) is the SAED pattern for the laser modified nanobelt. **(c** and **d)** HRTEM images of the nanobelts in (**a** and **b**), respectively.

TEM images and HRTEM images with SAED patterns of pristine and laser modified GeSe$_2$ nanobelts are shown in **Figure 6(a-d)**. TEM image (**Figure 6a**) and corresponding selected-area electron diffraction (SAED) pattern (inset) of the pristine GeSe$_2$ nanobelt indicate that smooth surfaced single crystalline nanobelts were employed for the laser pruning experiments. **Figure 6b** shows the TEM images and SAED pattern of the laser modified nanobelt. After laser modification, the surface of the nanobelt has been transformed into non-uniform amorphous like material. **Figure 6(c,d)** show the HRTEM images of the nanobelts as shown in **Figure 6(a,b)**, respectively. Lattice fringes of GeSe$_2$ crystal can be seen from HRTEM image (**Figure 6c**), whereas no lattice fringes were observed in laser modified HRTEM image (**Figure 6d**).

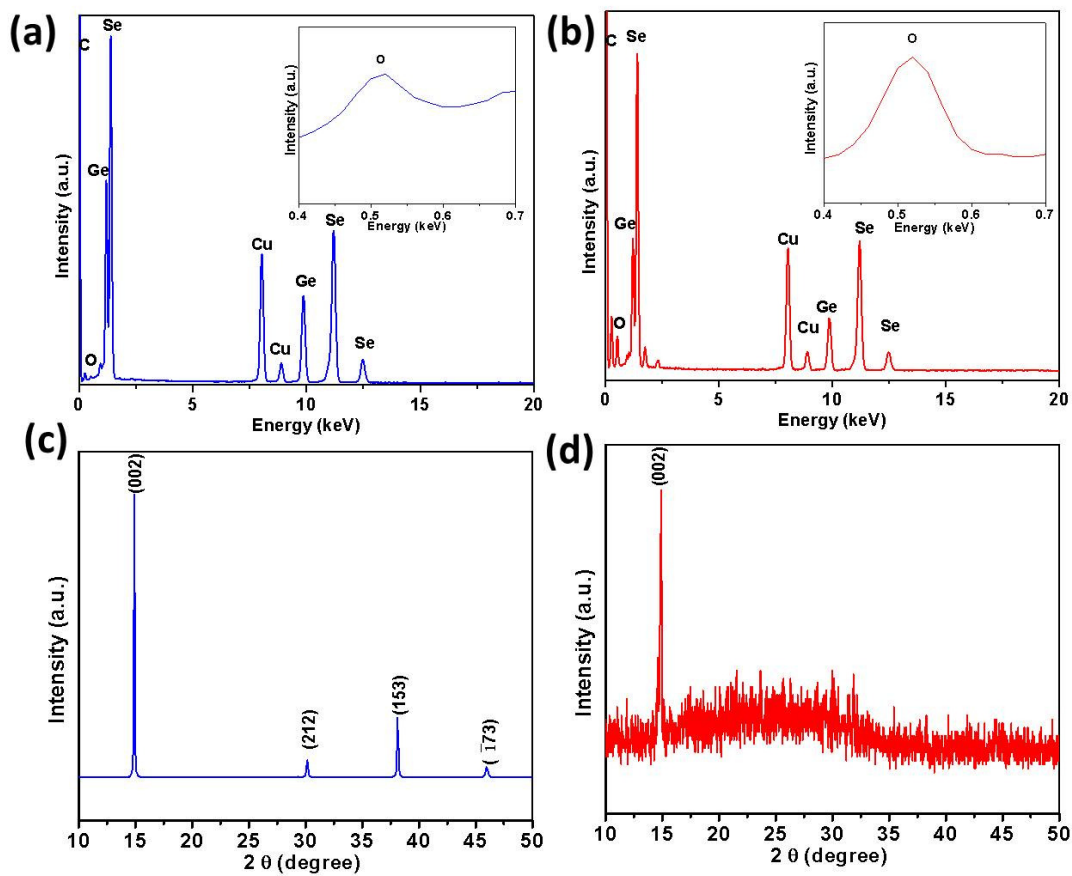

**Figure 7 (a), (b)** EDS spectra of GeSe$_2$ NSs before and after laser modification, respectively. Insets show the magnified EDS spectra of 'O' element for the respective curves. **(c)** and **(d)** XRD patterns of GeSe$_2$ NSs film before and after laser modification, respectively.

The photoinduced change in local structure is known to be present in GeSe$_2$ material.[36] The local irradiation of focused laser beam permanently modifies the local structure of NSs film, which results in permanent changes in color and structure of the nanomaterials. To investigate the laser induced changes in the chemical nature, elemental composition and crystallinity order of the GeSe$_2$ films, we prepared a large area of laser modified NSs film sample using ~2 mW focused laser in ambient conditions. Energy-dispersive X-ray spectroscopy (EDS) analysis has been carried out in the pristine and laser modified area. **Figures 7(a,b)** show the EDS spectrum taken from the pristine GeSe$_2$ NSs and laser modified NSs, respectively. The magnified EDS spectrum of oxygen, 'O', element for the respective curves are shown in the inset of the representative figures. Clearly higher atomic percentage of 'O' element is present in the laser modified NSs. Thus more oxide states related defect formation is obtained during laser modification of the sample. The atomic percentage ratio of "Se/Ge" is found to be ~ 2 and ~ 2.7 for pristine and laser pruned region, respectively. XRD studies were carried out before and after laser pruning NSs film sample. **Figures 7(c,d)** show the XRD patterns of the pristine nanostructure film and laser modified film, respectively. Intense diffraction peaks can be indexed to GeSe$_2$ crystal planes (Joint Committee for Powder Diffraction Studies, JCPDS, card no. 71-0117) for the NSs film before laser modification sample. No XRD peaks of GeSe, GeO$_2$, or other intense peak are detected. It is noted that after laser modification the crystalline quality has decreased drastically and a small background hump appeared in XRD pattern, which is due to the formation of amorphous material. This observation is consistent with TEM results at high energy laser modification.

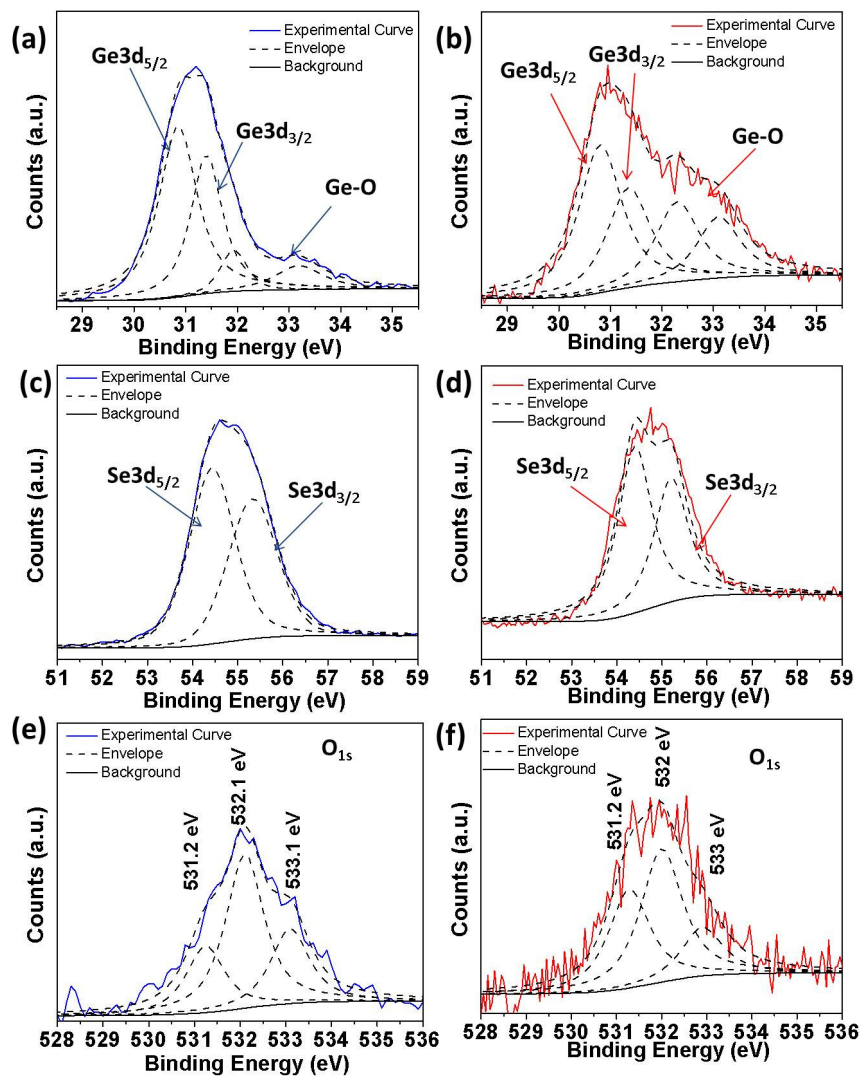

**Figure 8** XPS spectra of GeSe$_2$ NSs for pristine (left column) and pruned (right column) region: **(a),** and **(b)** for Ge element, respectively; **(c),** and **(d)** for Se element, respectively; and **(e),** and **(f)** for O element, respectively. A Shirley background correction was applied to each spectrum before curve fitting.

To further investigate the chemical changes during the laser pruning process, X-ray photoelectron spectroscopy (XPS) was carried out for pristine and laser modified regions as shown in **Figure 8**. Charge correction based on the standard reference signal from C 1s (284.8 eV) was applied to all XPS peaks in Ge 3d, Se 3d and O 1s obtained from pristine and laser treated samples. **Figure 8a** shows that Ge 3d core-level peaks for pristine sample centered at ~31.1 ± 0.2 eV, which coincides with the reported by low-resolution XPS for GeSe$_2$ (~31.1 eV) and the position is higher than that of pure Ge3d peak (~ 29 eV).[37,38] It is noted that Ge3d core-level spectrum (**Figure 8a**) has a very small shoulder on the high energy side, which could be due to the presence of Ge oxide (GeO$_x$) in the sample.[39-41] The shoulder peak position on the high energy side is O-related and has been ascribed to Ge-O bond states.[37,41] Thus oxide bond states formation is prominent in the laser modified NSs sample. **Figure 8b** shows Ge3d core-level peak at ~ 31.1 ± 0.2 eV for laser modified sample, which is at same binding energy with the peak position for pristine NSs sample. It has been noted that the peaks area under the core-level peak has greatly enhanced after laser modification of the NSs, which could be due to the modification of the morphology, chemical composition and crystallinity of the NSs. For laser modified sample, the peaks area of GeO$_x$ peaks (**Figure 8b**) has increased significantly, which indicates the presence of higher amount oxygen elements in the laser modified sample. **Figures 8(c,d)** show that Se 3d core-level peaks centered at ~54.7 ± 0.2 eV and ~54.7 ± 0.2 eV for pristine and laser modified samples, respectively, which is reported for GeSe$_2$ at ~ 54.7 eV.[38] There is no information about Se-O oxide bond state formation after the laser modification as seen from the Se 3d core level XPS peaks. The O1s peaks for pristine and laser modified NSs samples consist of various peaks as shown in **Figures 8(e,f)**. The asymmetric O1s peak for pristine NSs sample (**Figure 8e**) is resolved into three components at 531.2 ± 0.2, 532.1 ± 0.2 and 533.1 ± 0.2 eV. The peak at 531.2 ± 0.2 eV can be attributed to GeO$_x$ oxygen, whereas the shoulder peaks at 532.1 ± 0.2 eV and 533.1 ± 0.2 eV have been assigned to the chemisorbed oxygen and adsorbed water vapor, respectively.[42,43] The asymmetric O1s peak for laser modified GeSe$_2$ NSs sample (**Figure 8f**) is also resolved into three components at 531.2 ± 0.2, 532 ± 0.2 and 533 ± 0.2 eV. In the O1s peaks with the change in the peak area corresponding to pristine sample, which could be due to change of chemical compositions and due to more adsorbed oxygen species. The XPS results are also consistent with the EDS results.

To gain further insight into the role of environment oxygen in the modification by the focused laser beam, a specially designed vacuum chamber under controlled environment as shown in the schematic diagram of **Figure 1(a)** was used as a housing unit for the NSs film. Systematic studies of the dependency of the laser modification on gas environment have been carried out on the NSs film. We have used fixed laser power to modify the patterned regions with different environments. Three square boxes were patterned in air, vacuum and helium gas environment via the focused laser beam of 1.8 mW and the corresponding optical microscopy images of the patterned microsquares are shown in **Figure 9(a,b,c)**. We can see that the color change is more prominent and caustic in vacuum and helium compared to that in air. **Figure 9(d,e,f)** shows the SEM images corresponding to optical images **Figure 9(a,b,c)**. It is observed from SEM images that the difference in morphological changes was similar to laser modification under different gaseous environments, but the optical color contrasts have been changed significantly. Thus the gas molecules in air environment have great influence in changing the color contrast of laser modified region. In addition, Raman spectra (**Figure 9(g,h,i)**) show that the β-phase crystalline structure is changing in different gas environment. However the percentage of crystalline structure modification is different under different gaseous environments. Ratio of intensity of α crystalline ($I_\alpha$) to that of β crystalline ($I_\beta$) is ~ 1.3, ~ 0.6 and ~ 0.6 for the sample with laser modified in air, vacuum and helium, respectively.

Apart from the modification in the crystallinity, the laser pruning also results in the creation of different amount of surface states or defect states under different gas environments. Optical microscopy images (**Figure 9(a-c)**) show the distinct color changes of the microstructures modified in different gas environments, which could be due to different amount of elements present in them. Using TEM with EDS, we have obtained the EDS spectra of the modified microstructures in different gas environment, where the amount of elements presence in the region is shown in **Table 1**. The ratio of the atomic percentage (at%) of Se/Ge is ~ 1.7 for the sample modified in vacuum and helium environment, whereas the ratio turns to ~ 2.7 for the sample modified in air environment. There is also a higher percentage of oxygen element presence in the microstructure, which is modified in air atmosphere. Thus along with crystalline structure changes (from crystalline to amorphous like materials), the modification of chemical composition during laser modification is another contributing factor to the color changes,[35] where the environmental oxygen element has important role for such color changes.

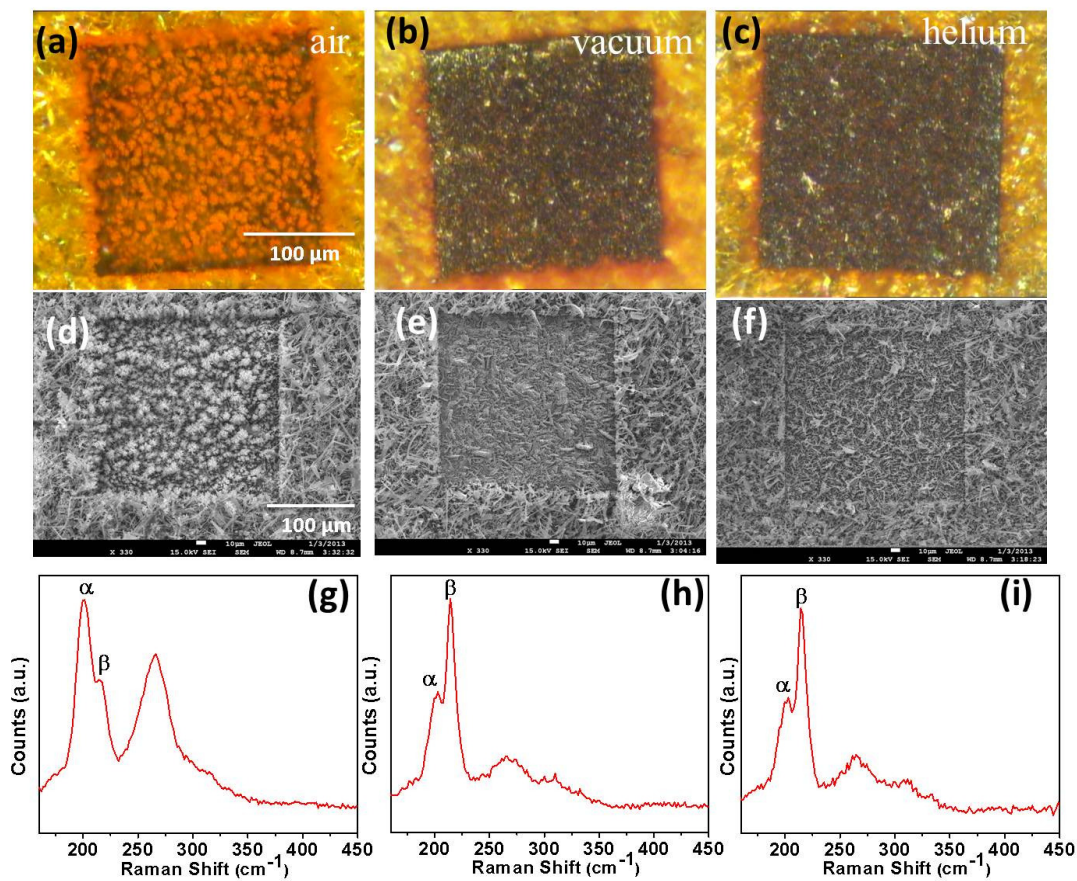

**Figure 9 (a,b,c)** Optical microscopy images of three microsquares patterns on as synthesized GeSe$_2$ NBs film after laser pruned (fixed laser power ~ 1.8 mW) in air, vacuum and helium environment, respectively. **(d,e,f)** SEM images corresponding to the optical images **(a,b,c)**. **(g,h,i)** Raman spectra of the sample with laser modified in air, vacuum and helium as shown in **Figure (a)**, **(b)** and **(c)**, respectively.

**Table 1**. Atomic Percentage of Ge, Se and O in NSs modified by focused laser beam under different environment.

| Laser modified NSs in | Average Atomic % of | | | Ratio Ge:Se:O |
|---|---|---|---|---|
| | Ge | Se | O | |
| Air | 18.8±2.0 | 51.6±5.2 | 29.6±5.5 | 1.0:2.7:1.6 |
| Vacuum | 33.9±0.7 | 60.6±1.0 | 5.3±1.6 | 1.0:1.7:0.1 |
| Helium | 34.9±1.0 | 60.6±1.2 | 4.4±2.3 | 1.0:1.7:0.1 |

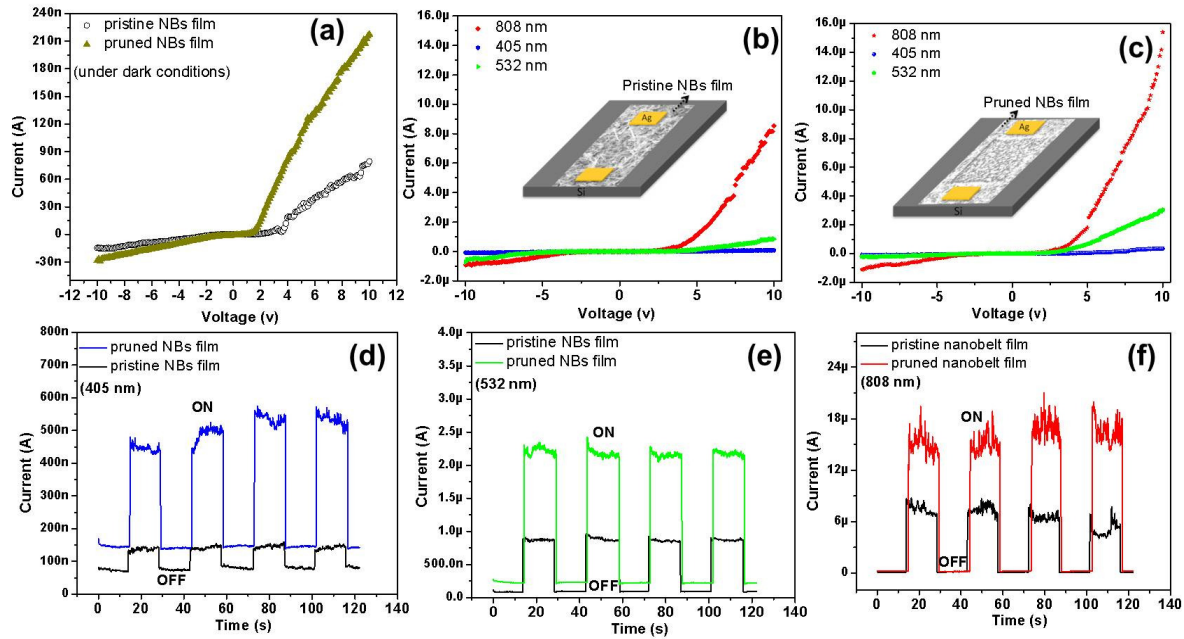

**Figure 10 (a)** I-V curves obtained from pristine GeSe$_2$ NSs film and after laser modification with two-probe measurements configuration. **(b)** I-V responses of the pristine NSs film under different laser sources with fixed laser intensity of ~ 0.8 mW/mm$^2$. **(c)** I-V responses of the same NSs film after laser modification with same experimental conditions. **(d)**, **(e)** and **(f)** I-t responses (fixed bias ~ 10V and fixed laser intensity ~ 0.8 mW/mm$^2$) of the NSs film before and after laser modification with laser wavelength of 405 nm, 532 nm, and 808 nm, respectively.

Besides the tuning of the structural properties, we have studied the optoelectrical properties of the laser pruned NSs. We made simple two probe devices by depositing silver paint directly on the pristine NSs film on Si substrate. The schematic diagram of the pristine nanostructure film device, which is fabricated in a very simple way, is shown in the inset of the **Figure 10(b)**. The laser modified nanostructure film placed in-between the electrical contacts of the device was used for the study of optoelectrical properties. The photoresponse measurements of pruned NSs are shown schematically in the inset of **Figure 10(c)**. The typical dark current-voltage (I-V) curves for the pristine and laser modified NSs films are shown in **Figure 10(a)**. It is noted from the dark I-V curves that the conductivity of the pruned NSs film has improved as compared to the pristine NSs film. At fixed positive bias of 10V, the film shows ~76 nA current whereas after modification of the nanostructure film, the dark current has improved to ~ 217 nA. Such an improvement can be attributed to the laser induced nanostructure modification, where the contacts between the nanobelts in the NSs network have been improved. Thus the results show the potential application for a simple one-step maskless, and in situ lithographic method for creating selective conduction regions. Beside the change in dark conductivity, the pristine NSs and laser pruned NSs has shown difference in photosensing properties when they are irradiated with laser sources of different wavelength. Note that in this case, we used a broad laser beam so the NSs were not modified by the laser irradiation. I-V responses of pristine GeSe$_2$ NSs film under broad beam irradiation of 405 nm, 532 nm and 808 nm light illumination and fixed laser intensity ~ 0.8 mW/mm$^2$ are shown in **Figure 10(b)**, whereas the I-V responses of the pruned NSs film under same laser light illumination are shown in **Figure 10(c)**. The results show that the pruned nanostructure film has higher photoinduced current than the pristine nanostructure film. The photocurrent-time (I-t) responses of the devices at 10 V fixed bias and fixed laser intensity ~ 0.8 mW/mm$^2$ under successive on/off operations of different laser sources have been compared. **Figure 10 (d-f)** shows the I-t response of the pruned and pristine NSs films under 405 nm, 532 nm and 808 nm light irradiation, respectively. The current for both pristine and pruned NSs films increased under broad beam laser with different wavelength illumination compared to that measured at a dark environment condition. From I-t curves it is observed clearly that the photoresponsivity is higher for pruned nanostructure film compared with pristine nanostructure film for all different laser wavelength irradiation.

It is known for GeSe$_2$ nanobelts that majority of the photoresponses of GeSe$_2$ nanobelt based devices come from the possible defects present in the materials.[20] After laser pruning, structural changes and more oxide states related defects are observed in the material. The higher photoresponsivity of the pruned nanostructure film when compared with pristine film could be attributed to the structural changes and introduction of more defects after laser pruning. These results demonstrate the enhanced optoelectrical properties of the laser pruned GeSe$_2$ NSs films and indicate their potential applications in optoelectronics such as photoswitching and wavelength selective photosensors.

**CONCLUSIONS**

In summary we have presented a simple technique to directly create multicolored microstructures on GeSe$_2$ NSs film. A set of well-defined microstructured boxes were patterned with different laser power. Controlled structural changes from β-crystalline to α-crystalline were observed with controlled laser modified nanostructured film. The effect of local heat production and focused laser induced modification of the NSs have been addressed. The proposed mechanism of laser induced crystalline structural deformation is the main driving force for local structural and chemical modification of the GeSe$_2$ NSs. In addition, laser modification under controlled gas environments shows the modification of chemical composition of the NSs, which causes different color contrast the optical microscopy images. The modification of chemical composition, crystalline structural modification and the gas molecules in air environment have great influence on changing the color contrast of laser modified region. Laser modified NSs film also exhibits better photoconducting properties.

## METHODS

**Synthesis of GeSe$_2$ NSs:** A mixture of pure Ge powder (Sigma Aldrich, purity 99.99%), Se powder (Sigma Aldrich, purity 99.99%) and carbon nanopowder (particle size < 50 nm, Sigma Aldrich, purity 99.99%) in molar ratio of 1:2:3, respectively was used as source materials. The Au coated (15 nm thick) Si substrate and the small alumina boat containing a small amount (~ 0.4g) of as milled Ge:Se:C source powders were loaded into an one-end-closed quartz-glass tube. The tube was inserted in a horizontal quartz tube placed in a conventional tube furnace such that the substrate was set at low-temperature region with respect to the mixed Ge:Se:C source powders and the distance between them was about 28 cm. Then the quartz tube was evacuated for 2 hrs by a vacuum pump and subsequently filled with Argon. The Argon gas was allowed to flow for 1 hr. After that, the furnace was heated under an Argon flow of 100 sccm (standard cubic centimeters per minute). When the temperature reached 680 $^{\circ}$C (heating rate: 30 $^{\circ}$Cmin$^{-1}$), the pressure of Ar carrier gas was maintained at ~ 2 mbar for 30 mins during synthesis. At the termination of the reaction, the substrate coated with a layer of yellow product was taken out.

**Characterization:** The morphology, structure and chemical composition of the as-synthesized NSs were characterized using field emission scanning electron microscopy (FESEM, JEOL JSM-6700F), energy-dispersive X-ray spectroscopy (EDX) equipped with SEM, transmission electron microscopy (TEM, JEOL, JEM-2010F, 200kV), X-ray photoelectron spectroscopy (XPS); Omicron EA125 analyzer using twin anode X-ray source (1253.6 eV), X-ray diffraction (X'PERT MPD, Cu Kα (1.5418 Å) and Raman spectroscopy (Renishaw system 2000, excitation 514.5 nm Ar$^+$ laser). All electrical and optoelectrical measurements were carried out using Keithley-6430 souse measurement unit under near vacuum condition at a pressure of 0.001 mbar. Laser beams were directly irradiated on the device through a transparent glass window of a vacuum chamber and the device was electrically connected directly through a pair of vacuum compatible leads to the source unit. Different continuous laser sources from the diode laser (405 nm, Power Technology, Inc.; 808 nm, EOIN and SUWTECH LDC-2500 diode laser of 532 nm) were used for the measurements of the photoresponse properties.

# Supporting Information

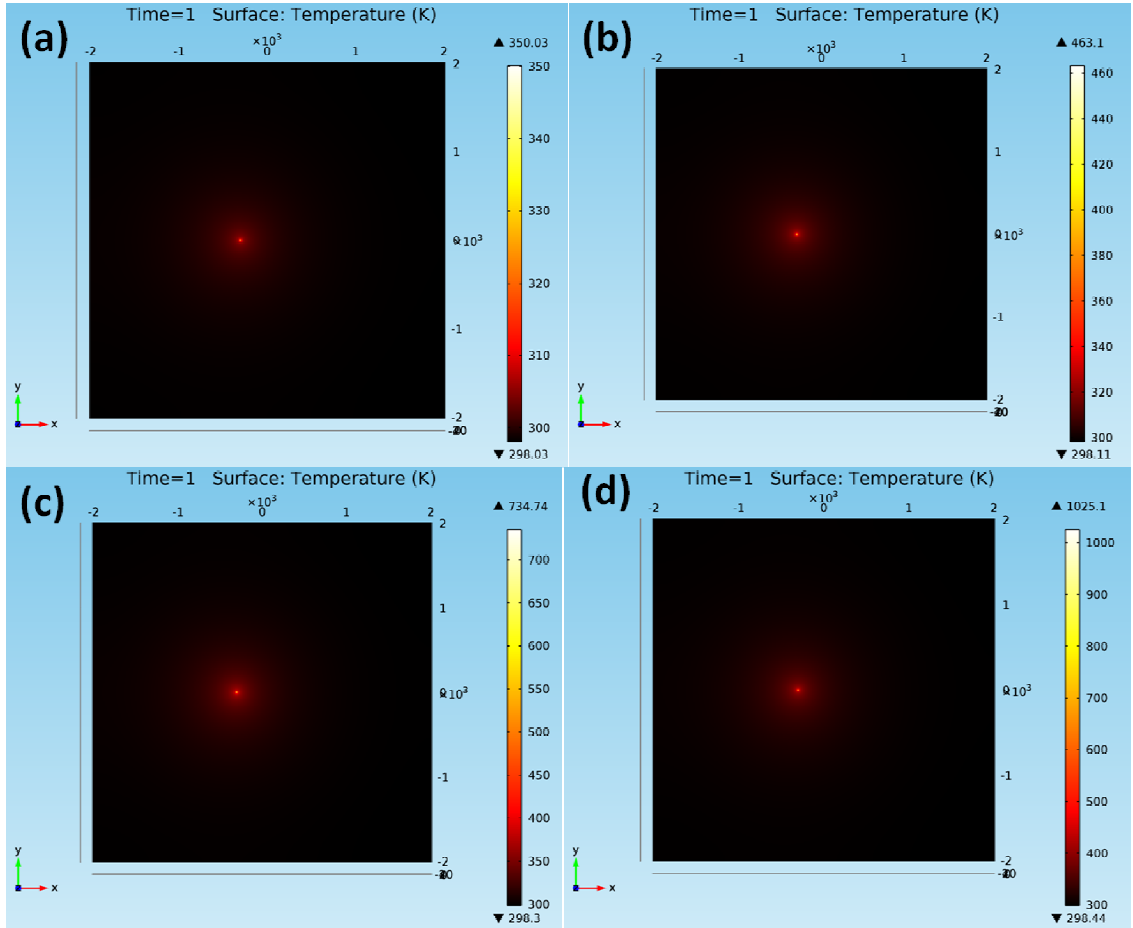

**Figure S1 (a), (b), (c) and (d)** XY-plane view of the temperature distribution on top surface of GeSe$_2$ NSs film surface with focused 532 nm laser beam (spot size ~ 3 μm) of different laser powers 0.2 mW, 0.6 mW, 1.8 mW and 40 mW, respectively.

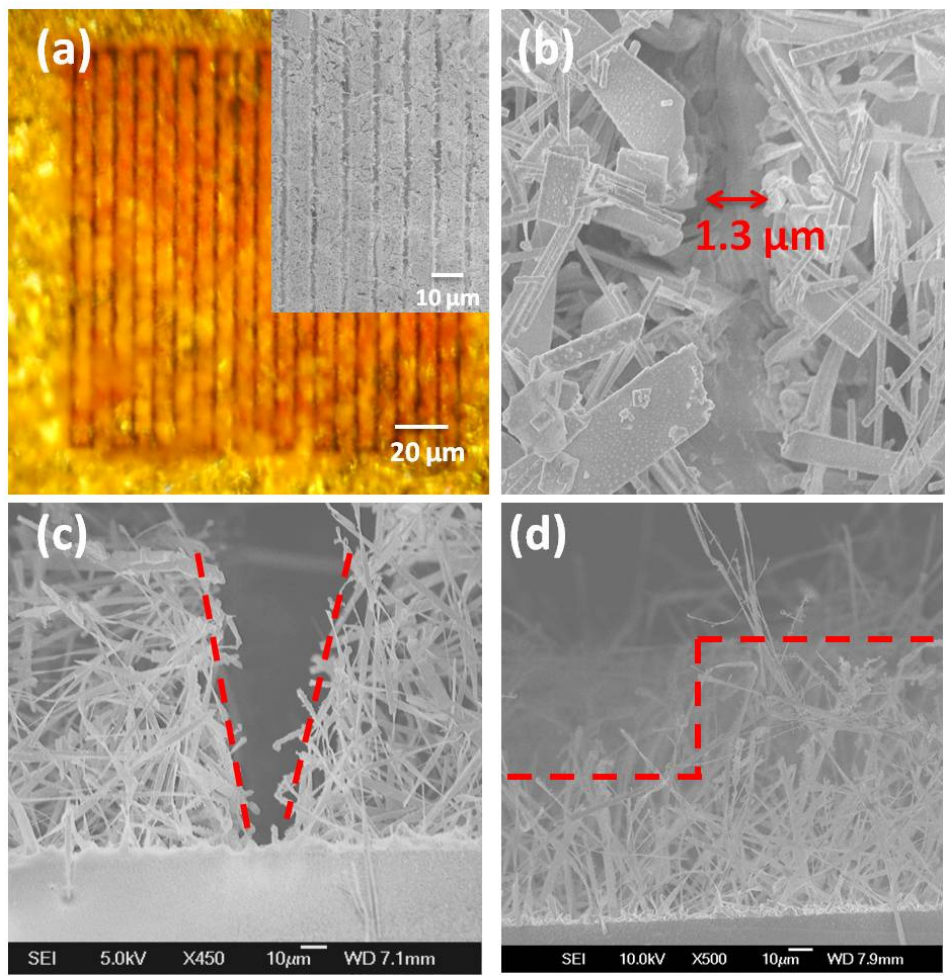

**Figure S2 (a)** Optical image of an array of microchannels. Inset shows the SEM image of the microchannels. **(b)** Magnified SEM image of 1.3 µm channel. **(c)** and **(d)** Cross sectional SEM images of a V shaped microchannel created using focused laser with high laser power ~ 40 mW and of the sample shown in **(d)**, respectively.